\documentclass[aps,pra,twocolumn,superscriptaddress]{revtex4-1}
\usepackage{graphicx}
\usepackage{amsmath}
\usepackage{subfigure}
\usepackage{braket}
\usepackage{color}
\begin{document}

%\preprint{}

\title{Retardation effects in spectroscopic measurements of the Casimir-Polder interaction }


\author{J. C. de Aquino Carvalho}
%\email{athanasios.laliotis@univ-paris13.fr}
\affiliation{Laboratoire de Physique des Lasers, Universit{\'e} Paris 13, Sorbonne Paris-Cit{\'e}, F-93430, Villetaneuse, France}
\affiliation{CNRS, UMR 7538, LPL, 99 Avenue J.-B. Cl{\'e}ment, F-93430 Villetaneuse, France}
\author{ P. Pedri}
%\email{athanasios.laliotis@univ-paris13.fr}
\affiliation{Laboratoire de Physique des Lasers, Universit{\'e} Paris 13, Sorbonne Paris-Cit{\'e}, F-93430, Villetaneuse, France}
\affiliation{CNRS, UMR 7538, LPL, 99 Avenue J.-B. Cl{\'e}ment, F-93430 Villetaneuse, France}
\author{ M. Ducloy}
%\email{athanasios.laliotis@univ-paris13.fr}
\affiliation{Laboratoire de Physique des Lasers, Universit{\'e} Paris 13, Sorbonne Paris-Cit{\'e}, F-93430, Villetaneuse, France}
\affiliation{CNRS, UMR 7538, LPL, 99 Avenue J.-B. Cl{\'e}ment, F-93430 Villetaneuse, France}
\affiliation{Division of Physics and Applied Physics, School of Physical and Mathematical Sciences, Nanyang Technological University, 637371 Singapore}
\author{ A. Laliotis}
\email{athanasios.laliotis@univ-paris13.fr}
\affiliation{Laboratoire de Physique des Lasers, Universit{\'e} Paris 13, Sorbonne Paris-Cit{\'e}, F-93430, Villetaneuse, France}
\affiliation{CNRS, UMR 7538, LPL, 99 Avenue J.-B. Cl{\'e}ment, F-93430 Villetaneuse, France}

%\thanks{}


\date{\today}

\begin{abstract}
Spectroscopy is a unique experimental tool for measuring the fundamental Casimir-Polder interaction between excited state atoms, or other polarisable quantum objects, and a macroscopic surface. Spectroscopic measurements probe atoms at nanometric distances away from the surface where QED retardation is usually  negligeable and the atom-surface interaction is proportional to the inverse cube of the separation distance, otherwise known as the van der Waals regime. Here we focus on selective reflection, one of the main spectroscopic probes of Casimir-Polder interactions. We calculate for the first time selective reflection spectra using the full, distance dependent, Casimir-Polder energy shift and linewidth. We demonstrate that retardation can have significant effects, in particular for experiments with low lying energy states. We also show that the effective probing depth of selective reflection spectroscopy depends on the transition linewidth.  Our analysis allows us to calculate selective reflection spectra with composite surfaces, such as metasurfaces, dielectric stacks, or even bi-dimensional materials.

\end{abstract}

%\keywords{out of equilibrium}

\maketitle
%\section{}
%\subsection{}
%\subsubsection{}
%\section{Introduction}

\section{Introduction}
The Casimir-Polder interaction of polarisable quantum objects, such as atoms or molecules, with a macroscopic surface is a fundamental problem of quantum electrodynamics. Spectroscopic measurement of atomic energy level shifts has  been one of the main experimental methods for probing atom-surface interactions. Spectroscopy of Rydberg atoms flying through metallic cavities was the first precision measurement of the van der Waals law \cite{hindsprl1992}, demonstrating that atom-surface potentials scale as $z^{-3}$, where z is the atom-surface separation. Selective reflection (SR), a technique used in conventional vapor cells, is also sensitive to atom surface interaction in the nanometric scale, probing atoms at distances on the order of $100\;\mathrm{nm}$ away from dielectric windows \cite{oriaepl1991,fichet_jphysique_1991}. Selective reflection has been used to demonstrate atom-surface repulsion of excited state atoms due to resonant coupling with surface polaritons \cite{failacheprl1999}, as well as to demonstrate a strong temperature dependence of the Casimir-Polder interaction due to thermal excitation of polariton modes \cite{laliotisnatcommun2014, passeratlaserphysics2014}. Thin cell transmission and reflection have been used to measure atom-surface interactions \cite{fichet_epl_2007} and more recently evidence of van der Waals interactions was also observed on thin cell fluorescence spectra \cite{PhysRevLett.112.253201}.

Spectroscopic probing of the Casimir-Polder interaction has been so far seemingly faithful to the $z^{-3}$, van der Waals law, whereas retardation effects, that were famously first predicted by Casimir and Polder \cite{casimirphysrev1948}, have  been demonstrated only with ground state atoms. This was either done by measuring the deflection of ground state sodium atoms \cite{hindscp1993}, or by using cold atom trapping in the vicinity of surfaces \cite{benderprl2010, harberpra2005, Landragin_prl_1996}. Nevertheless retardation effects for excited state atoms have  remained elusive in most spectroscopic experiments, with a notable exception of experiments performed with ions placed extremely far away (on the order of $20 \;\mathrm{cm}$) from a surface \cite{Blatt_prl_2004}. These experiments have shown QED oscillations of the Casimir-Polder force, similar to the ones predicted for a classical antenna, due to the influence of spontaneous emission. The intermediate regime of interaction has not been studied with excited state atoms. In fact, analysis of spectroscopic measurements, in particular selective reflection or thin cell spectroscopy, has only been performed under the prism of a pure van der Waals law \cite{fichet_jphysique_1991}. However, recent experimental and theoretical studies \cite{laliotis_pra_2015} suggest that retardation could have measurable effects for spectroscopic experiments with low-lying atomic energy states. 

Here we theoretically investigate the effects of Casimir-Polder retardation on selective reflection spectra. In section II we outline the principles of the calculation of selective reflection spectra accounting for a fully retarded Casimir-Polder potential. In section III, we calculate the Casimir-Polder potential of  Cesium low lying energy states and we present the theoretically predicted spectra of the corresponding selective reflection experiments. We show that retardation effects have an impact on predicted spectra and experimental measurements of the van der Waals coefficient.
Finally, section IV, we discuss how our analysis is imperative for interpreting spectroscopic measurements with more complex geometries such as  meta-surfaces that now offer an attractive way for tuning the Casimir-Polder interaction via tuning of surface plasmon or polariton resonances \cite{aljunid_atomic_2016}. Our approach allows us to account for a distance dependent shift and linewidth. This can be important in the quest for identifying more delicate effects such as quantum friction \cite{klatt_pra_2016} in spectroscopic experiments.

\section{Influence of Casimir-Polder interaction on selective reflection spectrum}
The Casimir-Polder interaction has been theoretically investigated in numerous studies. Here, we follow the formalism introduced by Wylie and Sipe \cite{WSpra1984,WSpra1985}, since our emphasis will be on excited state atoms. The same formalism has been used to analyze a temperature dependent  Casimir-Polder interaction \cite{gorzaEPJD2006} and later to demonstrate that the temperature dependent Casimir-Polder interaction is equivalent to a shift induced by near field thermal emission \cite{laliotis_pra_2015}. 

For a given atomic state $ \ket{a}$ the free energy shift $\delta F_{a}$ due to the atom-surface interaction can be expressed as the sum of contributions resulting from all dipole allowed couplings, $\delta F_{a \rightarrow b}$, $\left( \delta F_{a}=\sum\limits_{b}\delta F_{a \rightarrow b}\right)$ 
which can in turn be decomposed in a resonant $\delta F^{r}_{a \rightarrow b}$ and non-resonant $\delta F^{nr}_{a \rightarrow b}$ contribution. The resonant term of the interaction is reminiscent of a classical interaction between an oscillating dipole and its image \cite{hindspra1991}. The non-resonant term originates from the QED picture of an atom  interacting with the fluctuating vacuum at non-zero temperature \cite{WSpra1985,hindspra1991, gorzaEPJD2006}. It can be viewed as a distance dependent Lamb shift \cite{WSpra1984, WSpra1985, gorzaEPJD2006}. The resonant and non-resonant terms of the Casimir-Polder interaction are given by the following expressions:
 
\begin{equation}
\delta F_{a \rightarrow b}^{nr} =-2\frac{k_{B} T}{\hbar}  {\sum_{p=0}^{\infty}}^{\prime} \mu_{\alpha}^{ab}\mu_{\beta}^{ba}G_{\alpha\beta}(z,i\xi_{p}) \frac{\omega_{ab}}{\xi_{p}^2+\omega_{ab}^{2}}
\label{eqn1}
\end{equation}

\begin{equation}
\delta F_{a \rightarrow b}^{r} =n(\omega_{ab},T) \mu_{\alpha}^{ab}\mu_{\beta}^{ba} Re\left[ G_{\alpha\beta}(z,|\omega_{ab}|)\right]
\label{eqn2}
\end{equation}
Here, $\omega_{ab}$ is the transition frequency that can be either positive or negative depending on the coupling, $\xi_{p}=2 \pi \frac{k_{B} T}{\hbar} p$ are the Matsubara frequencies and $n(\omega_{ab},T)$ is the Bose-Einstein factor. The prime symbol signifies that the first term of the sum should be multiplied by 1/2. We use the Einstein notation, implying a summation over the index variables $\alpha$ and $\beta$ that denote the Cartesian coordinate components. Finally, $ \mu_{\alpha}^{ab}$ and $\mu_{\beta}^{ba}$ are the dipole moment matrix elements and  $G_{\alpha\beta}(z,i\xi_{p})$ are the components of the linear susceptibility matrix of the reflected field defined in \cite{WSpra1984, WSpra1985}. The linear susceptibility matrix gives the reflected displacement field at a point $ \vec{r}$ due to a dipole  $\vec{\mu}(\omega)$, oscillating at a frequency $\omega$, positioned at $ \vec{r}\,'$, via the relation $\vec{D}( \vec{r}, \vec{r}\,',\omega)=\stackrel{\leftrightarrow}{G}( \vec{r},\vec{r}\,', \omega) \vec{\mu}(\omega) $. In our case $\stackrel{\leftrightarrow}{G}$  is evaluated for $ \vec{r}=\vec{r}\,'$, because we're interested in dipoles  interacting with their own reflected field. Due to the cylindrical symmetry $\stackrel{\leftrightarrow}{G}$ is only a function of frequency and distance $ z $ of the dipole from the reflecting wall. More details on the calculation of the elements of the linear susceptibility matrix are given in \cite{WSpra1984, WSpra1985,gorzaEPJD2006, laliotis_pra_2015}.

The distance dependent linewidth, $\delta \gamma_a(z)$, is also a summation  of contributions, $\delta \gamma_{a \rightarrow b}$, given by:

\begin{equation}
\delta \gamma_{a \rightarrow b} =2 n(\omega_{ab},T) \mu_{\alpha}^{ab}\mu_{\beta}^{ba} \mathrm{Im}\left[ G_{\alpha\beta}(z,|\omega_{ab}|)\right]
\label{eqn3}
\end{equation}

The far field limit ($z \gg \frac{\lambda_{ab}}{4 \pi}$) of the free energy shift $\delta F_{a \rightarrow b}$ and linewidth $\delta \gamma_{a \rightarrow b}$ are given by :

\begin{equation}
\delta F_{a \rightarrow b} =n(\omega_{ab},T) \mu_{\alpha}^{ab}\mu_{\beta}^{ba} \frac{k_{ab}^2}{z}|r(\omega_{ab})|\cos(2k_{ab} z+\phi(\omega_{ab}))
\label{eqn4}
\end{equation}

\begin{equation}
\delta \gamma_{a \rightarrow b} =2 n(\omega_{ab},T) \mu_{\alpha}^{ab}\mu_{\beta}^{ba} \frac{k_{ab}^2}{z}|r(\omega_{ab})|\sin(2k_{ab} z+\phi(\omega_{ab}))
\label{eqn5}
\end{equation}

where $\lambda_{ab}$, $k_{ab}$ are the transition wavelength and wavevector, and $r(\omega_{ab})=|r(\omega_{ab})|e^{i\phi(\omega_{ab})}$ is the surface reflection coefficient. In the near field ($z\ll\frac{\lambda_{ab}}{4 \pi}$), the free energy shift follows the well known van der Waals law that writes $\delta F_{a \rightarrow b}=-\frac{\mathrm{Re}[C_{3}]}{z^3}$. In the case of a dissipative surface (non-zero imaginary part of the dielectric constant), the distance dependent linewidth also follows the inverse cube law: $\delta \gamma_{a \rightarrow b}=-\frac{2 \mathrm{Im}[C_{3}]}{ z^3}$, where $C_{3}$ is the complex van der Waals coefficient.

Using the above definitions we can proceed to the calculation of the selective reflection spectrum using a fully retarded Casimir-Polder shift and linewidth. Selective reflection is a linear spectroscopic technique  that measures the reflection of a laser beam, near resonant with an atomic transition, at the interface of an atomic vapor and a dielectric surface (transparent at the laser frequency). Due to collisions with the dielectric surface the interaction of the atoms with the laser field is interrupted. As such a correct description of selective reflection takes into account the transient regime of atom-laser interaction \cite{fichet_jphysique_1991,SautenkovSR1982}. In its FM (Frequency Modulation) version selective reflection is linear (with respect to laser power), has a sub-Doppler resolution and is essentially sensitive to atoms that are at distances on the order of $\lambda/2\pi$ away from the dielectric surface, where $\lambda$ is the wavelength of optical excitation. The combination of high frequency resolution and detection of atoms at nanometric distances from the surface makes selective reflection a major experimental method for probing Casimir-Polder interactions of excited state atoms. Additionally selective reflection has been used for measuring the collisional broadening (broadening due to inter-atomic collisions) of atomic transitions \cite{papageorgiou_laserphyics_1994,SautenkovSR1982}. The possibility of measuring local-field corrections (Lorentz-Lorenz shift) at high vapor densities with strong laser attenuation inside the atomic vapor has also been considered \cite{GuoPRA1996}.

In  our study we usually consider transitions between the fundamental electronic state of the atom $|g\rangle$ and an excited state $|e\rangle$. The details of the calculation have been outlined in \cite{fichet_jphysique_1991}. Here we briefly remind that the calculation considers the transient atomic response to correctly describe the effective linear susceptibility of the atomic vapour. When a frequency modulation (FM) is applied to the laser probe beam the observable signal is in fact the derivative of the reflectivity as a function of laser frequency $\omega$ which according to \cite{fichet_jphysique_1991} is given by the following formula: 
\begin{equation}
S_{FM}={\rm Im}\left[ \int_{0}^{\infty}dz \int_{0}^{\infty}dz^\prime \frac{(z^\prime-z) e^{i k(z+z^\prime)}}{L(z^\prime)-L(z)} \right]
\label{eq6}
\end{equation}
The integral inside the brackets will be denoted as $I$. Here, $k$ is the laser wavevector and $L$ defined as the following indefinite integral: 
\begin{equation}
L(z)=\int \left[ \frac{\gamma+\delta \gamma(z)}{2}-i\left( \omega-\omega_{o}-\delta F(z) \right)  \right] dz
\label{eq7}
\end{equation}
where $\omega$, $\omega_o$ are the laser and transition frequencies and $\delta F(z)=\delta F_{e}(z)-\delta F_{g}(z)$  is the difference between the free energies between the probed states, which is the relevant quantity in selective reflection spectroscopy. $\gamma $ is the transition linewidth in the volume (away from the surface),  defined as the natural linewidth plus any additional collisional broadening,  and $\delta \gamma(z)=\delta \gamma_{e}(z)+\delta \gamma_{g}(z)$ is the distant dependent transition linewidth that essentially contains all surface effects. We can also write Eq. (\ref{eq7}) as:  

\begin{equation}
L(z)=L_{o}z-i \xi(z)
\label{eq8}
\end{equation}
where $L_{o}=\frac{\gamma}{2}-i(\omega-\omega_{o})$. The effects of the surface on the atomic properties are essentially contained in the indefinite integral $\xi(z)=  \int \left( -\delta F(z)+i \frac{\delta \gamma (z)}{2} \right) dz$. Here the shift $-\delta F(z)$ and the linewidth 
$\frac{\delta \gamma (z)}{2}$ appear as a real and imaginary parts of the Casimir-Polder potential respectively. The integration constant has been omitted as we are only interested in the difference$L(z^\prime)-L(z)$.
 
In most spectroscopic experiments one fits the experimental data with a theoretical model to extract information about the Casimir-Polder interaction \cite{hindsprl1992,oriaepl1991,failacheprl1999,laliotisnatcommun2014,
passeratlaserphysics2014, fichet_epl_2007, PhysRevLett.112.253201}. Here, however, $\xi(z)$ is a numerically calculated function that uses the theoretically estimated Casimir-Polder potential, without accounting for any adjustable parameters. For this purpose we rewrite Eq. (\ref{eq8}) using a dimensionless multiplicative constant, $\eta$, that is applied equally to both the free energy shift and linewidth : 
\begin{equation}
L(z)=L_{o}z-i\eta \xi(z)
\label{eq9}
\end{equation} 
$\eta$ changes the strength of the potential, and provides an adjustable parameter that can be used to fit the theoretical model to experimental data. For the purposes of this manuscript, selective reflection spectra are calculated using strictly the theoretical predictions for atom-surface potential (i.e $\eta=1$).

After a change of variables and some tedious algebra the selective reflection integral is written as:

\begin{equation}\begin{split}
I=\frac{2}{(1-i \Delta) \gamma_{o} k^2} \left[ \frac{1}{(-i+\alpha)^2} \right]+\frac{2}{(1-i \Delta) \gamma_{o} k^2} \times
 \\ 
 \left[ \int_{0}^{\infty}ds e^{is}e^{-\alpha s}  \int_{0}^{s}dt \frac{i \mathcal{A} \Xi(s,t)}{(1-i \Delta) -i \mathcal{A} \Xi(s,t)} \right]
\label{eq10}
\end{split}\end{equation}

The details of the calculation and the definition of $\Xi(s,t)$ are given in appendix A (see Eq. (\ref{eqA3})). Here we define the normalized frequency (detuning parameter) $\Delta=\frac{2(\omega-\omega_{o})}{\gamma}$ and the parameter $\mathcal{A} = \frac{2 \eta k}{\gamma}  $. The parameter $\alpha$ is the attenuation coefficient due to the exponential laser absorption inside the resonant vapor (see also \cite{fichet_jphysique_1991,GuoPRA1996}). In normalized frequency units the shape of the spectrum depends exclusively on the parameter $\mathcal{A} $, which is essentially the ratio of the strength of the potential $\eta$ over $\gamma$, that defines the resolution of the experiment.

The result of Eq. (\ref{eq10}) displays many similarities with the selective reflection spectrum assuming a pure van der Waals potential \cite{fichet_jphysique_1991}. However, here the curves are not universal, since the distance dependence of the potential depends on the probed transition. Additionally, the calculation of the integrals is significantly more difficult. 
  
It is also worth mentioning that, while in the near field the distance dependent linewidth is usually significantly smaller than the atomic energy shift and in most cases can be safely ignored, this is not the case when one considers the complete Casimir Polder potential. It can be seen from Eq.(\ref{eqn4},\ref{eqn5}) that in the far-field, linewidth ($\delta \gamma/2$) and shift ($\delta F$) oscillate with the same amplitude and frequency and a phase shift of $\pi/2$. As such, ignoring the distance dependent linewidth in a fully retarded calculation has no realistic justification and can lead to erroneous results or even, in some cases, to divergent selective reflection integrals.

\section{RESULTS}

We now turn our attention to some specific cases, focusing mainly on Cesium which is widely used in spectroscopic SR experiments (see for example \cite{laliotisnatcommun2014, passeratlaserphysics2014, aljunid_atomic_2016,failacheEPJD2003} and references therein). In particular we examine the low lying excited states where dipole moment fluctuations remain relatively small, and comparable to those of the ground state. Here we also take  into account the modification of the spontaneous emission rate near the surface \cite{Lukosz_josa_1977} due to the reflection of the emitted field on the surface, or due to emission in the forbiden cone of the dielectric \cite{Burgmans_pra_1977} and in evanescent plasmon-polariton modes.  Within the near-field approximation the transition linewidth can be written as: $\delta \gamma_{a \rightarrow b} \propto  \frac{1}{z^3} \mathrm{Im}\left[\frac{\epsilon(|\omega_{ab}|)-1}{\epsilon(|\omega_{ab}|)+1} \right]$, which diverges close to the surface, if the surface dissipation is non-zero at the transition frequency ($\mathrm{Im}(\epsilon(|\omega_{ab}|)\neq 0 $). This represents an increase in the spontaneous emission rate of the atom due to the existence of evanescent, plasmon-polariton type, modes \cite{failache_prl_2002}. In the cases considered here this contribution is small for experimentally meaningful distances. Therefore, we consider the surface dissipation equal to zero ($\mathrm{Im}(\epsilon(|\omega_{ab}|)=0$), thus avoiding any divergence of the atomic linewidth very close to the surface (see also relevant discussion in \cite{WSpra1984}).
\begin{figure}[!htb]
\includegraphics[width=90mm]{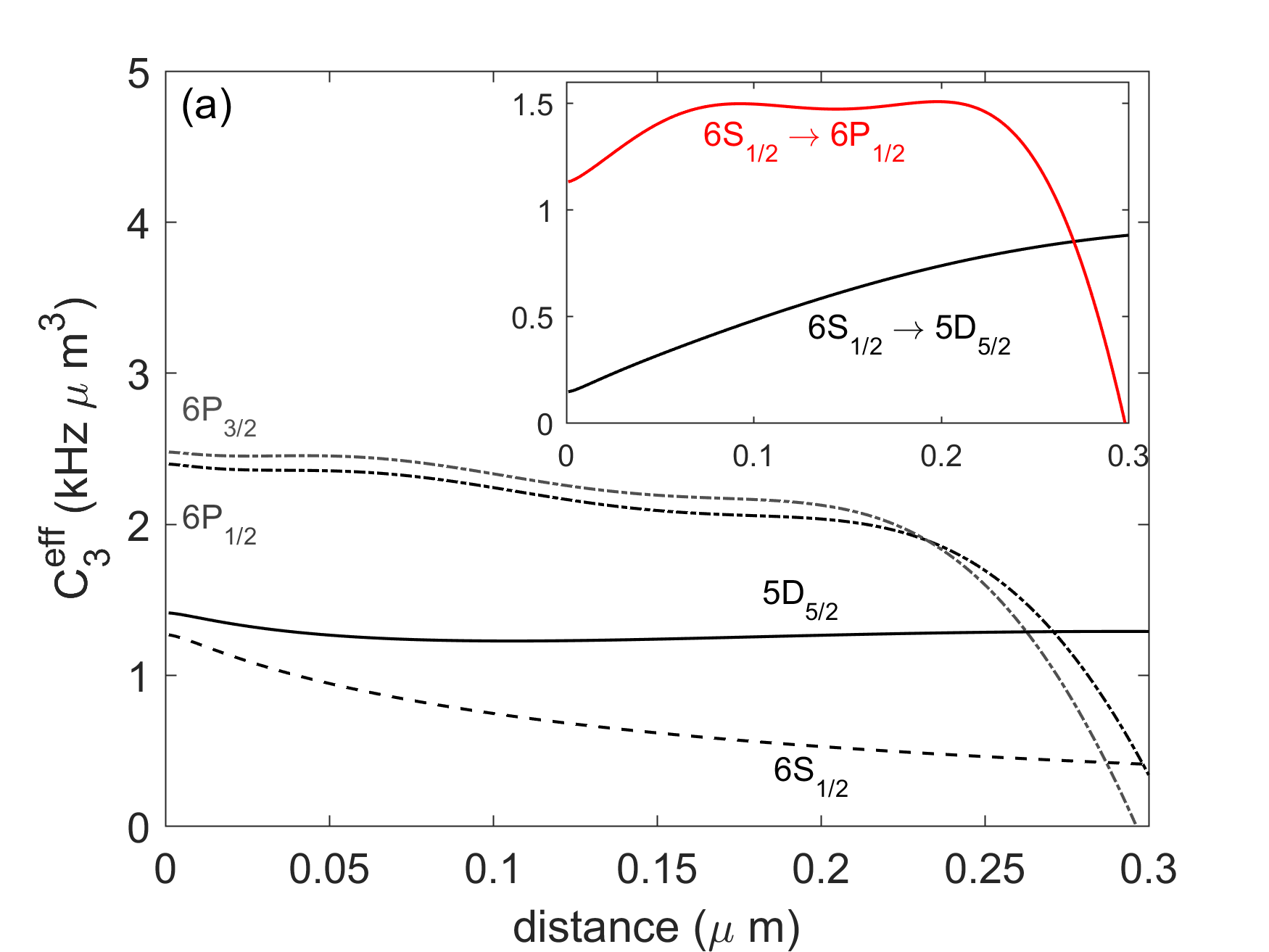}\\%
\includegraphics[width=90mm]{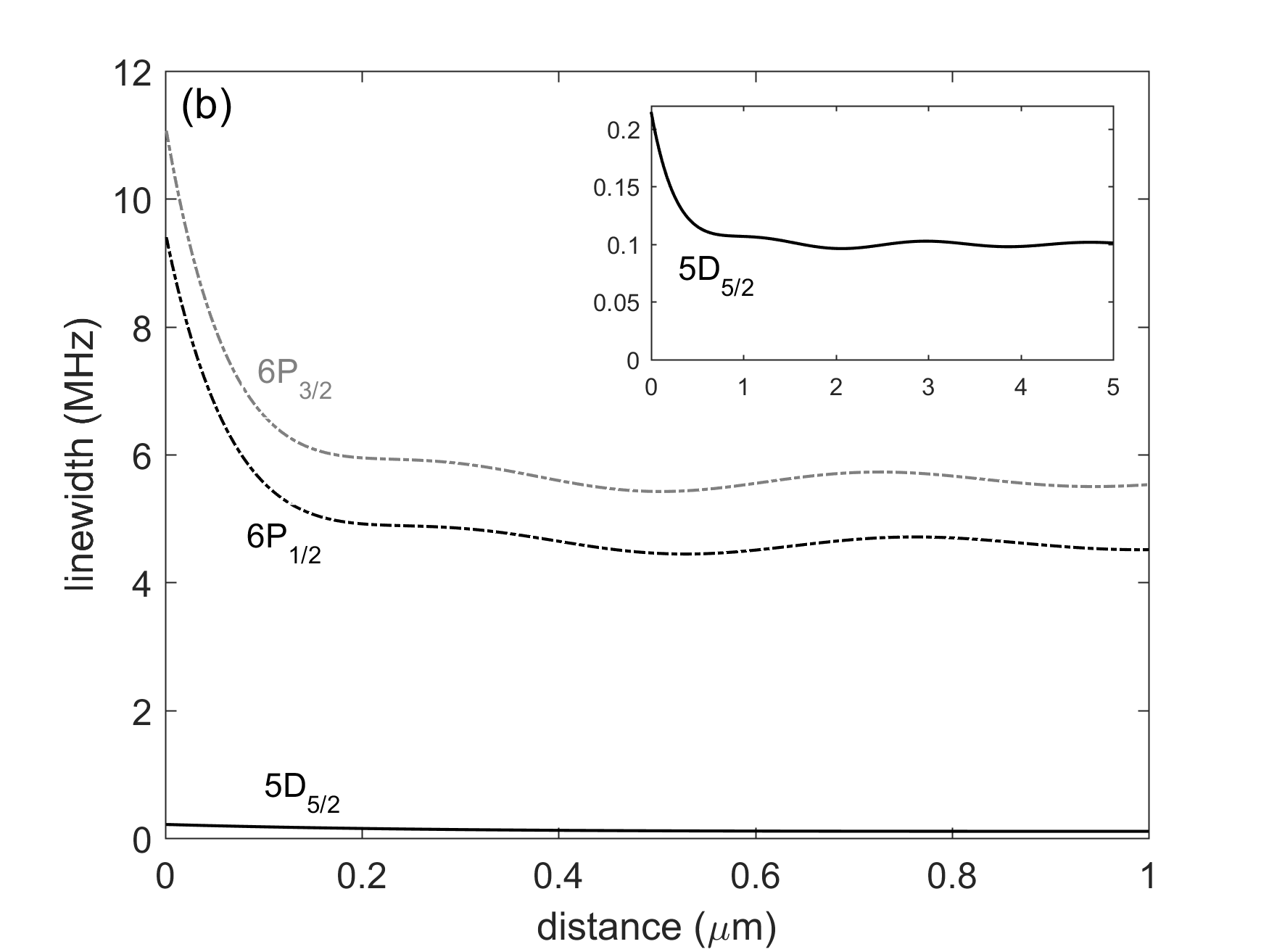}%
\caption{ (a) Effective  van der Waals coefficient $C^{\mathrm{eff}}_{3}$ (as defined in the main text)for the Cesium  levels $6S_{1/2}$ (black dashed line), $6P_{1/2}$ and $6P_{3/2}$ (black and grey dash-dotted lines respectively) as well as $5D_{5/2}$ (black solid line) against a sapphire surface. The inset shows the difference of the effective van der Waals coefficients for the  $6S_{1/2}\rightarrow5D_{5/2}$ (black solid line) and the  $6S_{1/2}\rightarrow6P_{1/2}$ (red solid line) transitions. (b) Distance dependent  linewidth for three principal transitions $6S_{1/2}\rightarrow 6P_{1/2}$ (black dash-dotted line), $6S_{1/2} \rightarrow 6P_{3/2}$ (grey dash-dotted line), $6S_{1/2}\rightarrow 5D_{5/2}$ (black solid line).
\label{Fig1}}
\end{figure}

In Fig.\ref{Fig1}(a) we show the energy level shifts for the Cesium levels  $6S_{1/2}$, $6P_{1/2}$, $6P_{3/2}$, and $5D_{5/2}$ against a sapphire surface, multiplied by the cube of the atom-surface distance $z$ ($- \delta F(z)\, z^3$). For simplicity, we call the quantity  $- \delta F(z)\, z^3$ an effective  van der Waals coefficient $C^{\mathrm{eff}}_{3}$. Our calculation is performed for a sapphire surface whose dielectric constant is given in \cite{barkerPR1963, passeratJP2009}. From Fig.\ref{Fig1}(a) we can see that for the excited states of Cesium, the $C^{\mathrm{eff}}_{3}(z)$ is practically constant within a few hundred of nanometers from the surface, whereas the ground state of Cesium  $6S_{1/2}$ decays much more rapidly, towards an  asymptotic $z^{-4}$ regime. This is partly because excited states present many dipole couplings at near and mid-infrared wavelengths but also because these couplings are both in absorption (positive transition frequencies) as well as in emission (negative transition frequencies). As such, excited states are sensitive to the distance dependence of the resonant term of the Casimir-Polder interaction whose distance dependence is very different from that of the non-resonant term.

We will examine here in more detail the spectra of selective reflection at the $6S_{1/2}\rightarrow6P_{1/2}$ and $6S_{1/2}\rightarrow5D_{5/2}$ transitions. The difference of the effective van der Waals coefficients for these experiments, representing the spectroscopically relevant quantity is shown as an inset of Fig.\ref{Fig1}(a). The first transition, the $\mathrm{D}_{1}$ line of Cesium, was already investigated experimentally , albeit with a cell containing significant quantities of buffer gas impurities \cite{laliotisAPB2008}. The $\mathrm{D}_{2}$ line of cesium ( $6S_{1/2}\rightarrow6P_{3/2}$ ), experimentally investigated in \cite{oriaepl1991, papageorgiou_laserphyics_1994}, exhibits a very similar behavior to the $\mathrm{D}_{1}$ line.

\begin{figure}[!htb]
\includegraphics[width=90mm]{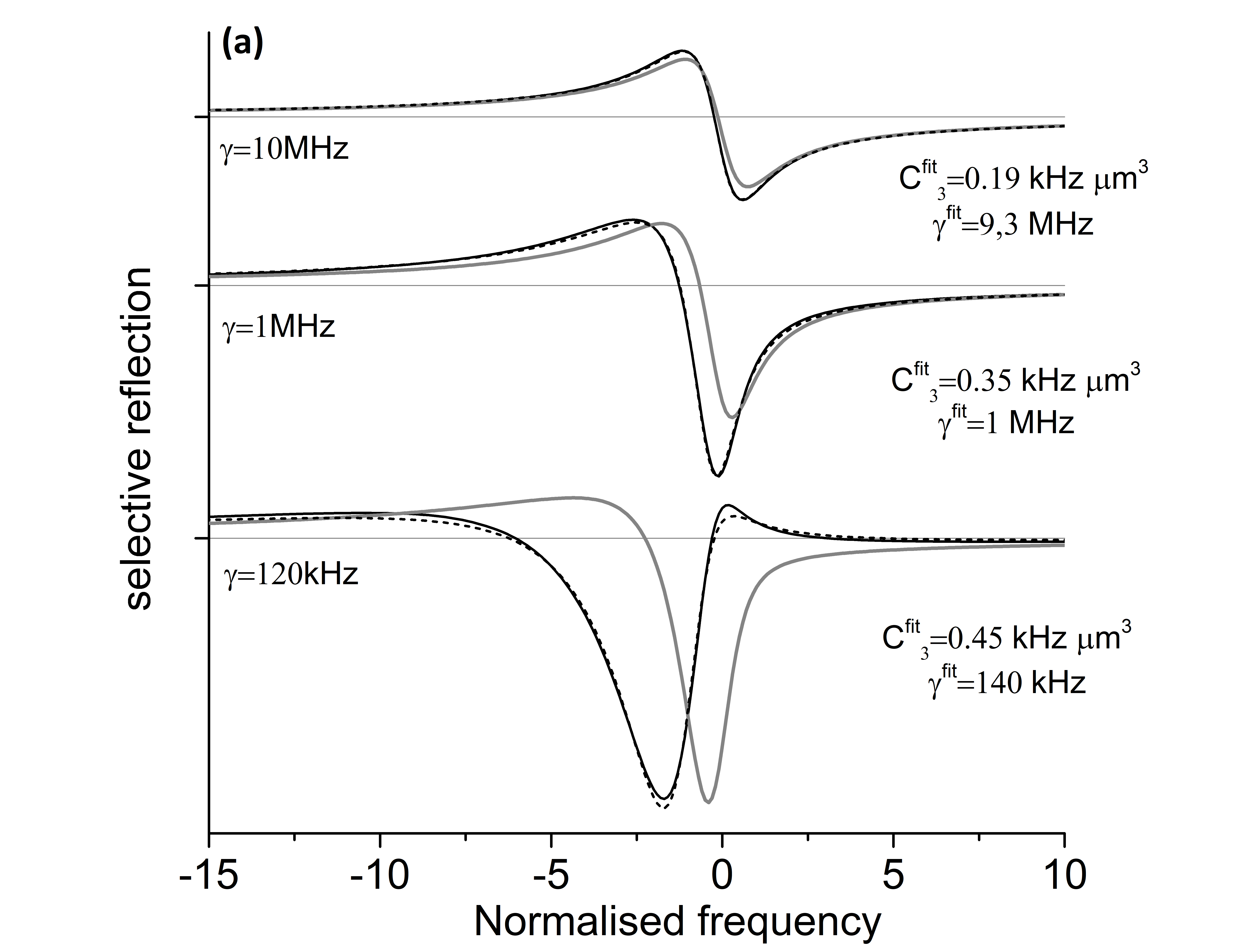}\\%
\includegraphics[width=85mm]{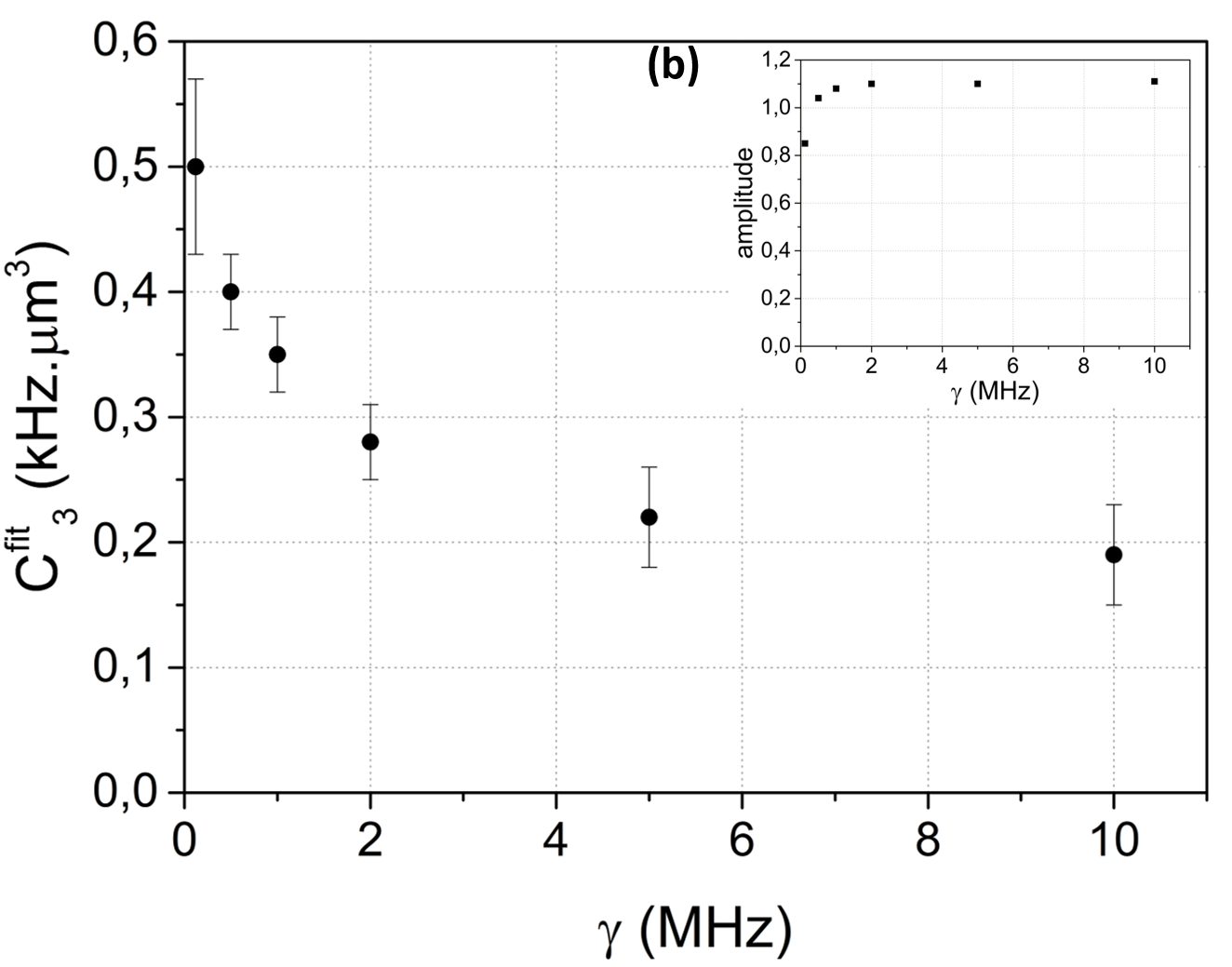}%
\caption{ (a) The black lines represent  the simulated selective reflection spectra ($S_{FM} $) on the $6S_{1/2} \rightarrow 5D_{5/2}$ transition, using a fully retarded Casimir-Polder potential (exact SR lineshapes) with a transition linewidth of $\gamma=120\;\mathrm{kHz}$ (natural linewidth) as well as $\gamma=1\; \mathrm{MHz}$ and $\gamma=10\;\mathrm{MHz}$ (assuming a collisional broadening). The spectra are given as a function of the normalized frequency $\Delta$, as defined in the text. The grey lines represent the expected SR lineshapes assuming a pure van der Waals atom-surface potential (i.e using the theoretical prediction of $C_{3}=0.15\;\mathrm{kHz}\,\mu \mathrm{m}^3$). The dashed curves are the best fits of the exact SR lineshapes using an ad hoc van der Waals  coefficient $C_{3}^{\mathrm{fit}}$. (b) The ad hoc van der Waals coefficient $C_{3}^{\mathrm{fit}}$ as a function of the transition linewidth. As the transition linewidth increases SR is more sensitive to atoms that are close to the surface and the values of $C_{3}^{\mathrm{fit}}$ approach the theoretical estimate of the van der Waals coefficient (Fig.\ref{Fig1}). The inset shows the amplitude ratio between the fully retarded SR spectra and the corresponding fits.
\label{Fig2}}
\end{figure}

In the case of the $6S_{1/2}\rightarrow5D_{5/2}$ transition SR is almost exclusively sensitive to retardation  mostly because the van der Waals coefficients of the two levels are very similar in magnitude. The $6S_{1/2}\rightarrow5D_{5/2}
$ transition is an electric quadrupole coupling, with small transition probability. Nevertheless, it has been experimentally probed by reflection spectroscopy of evanescent waves \cite{tojo_prl_2004} and more lately with high resolution pump-probe spectroscopy \cite{chan_optlett_2016}. Additionaly the    $5D_{5/2}$ level can be reached with a two photon, or Raman-type transition using two excitation lasers and appropriatly large detuning to minimise the influence of the intermediate state. Therefore the analysis that we will present here is much more than a simple theoretical curiosity. In Fig.\ref{Fig1}(b) we show the distance dependent linewidths for the $6P_{1/2}$ and $5D_{5/2}$ transitions (starting from the cesium ground state). The increase in linewidth (decrease in lifetime) observed close to the wall is a well known effect that depends on the orientation of the atomic dipole \cite{Lukosz_josa_1977,WSpra1984}, which is here considered to be random. A few hundreds of nanometers away from the surface, we observe QED oscillations of the linewidth (see Eq. (\ref{eqn5})) around its asymptotic value, which, for the purposes of Fig.\ref{Fig1}, is considered to be equal to the natural transition linewidth assuming zero collisional broadening.

\begin{figure}[!htb]
\subfigure{\includegraphics[width=90mm]{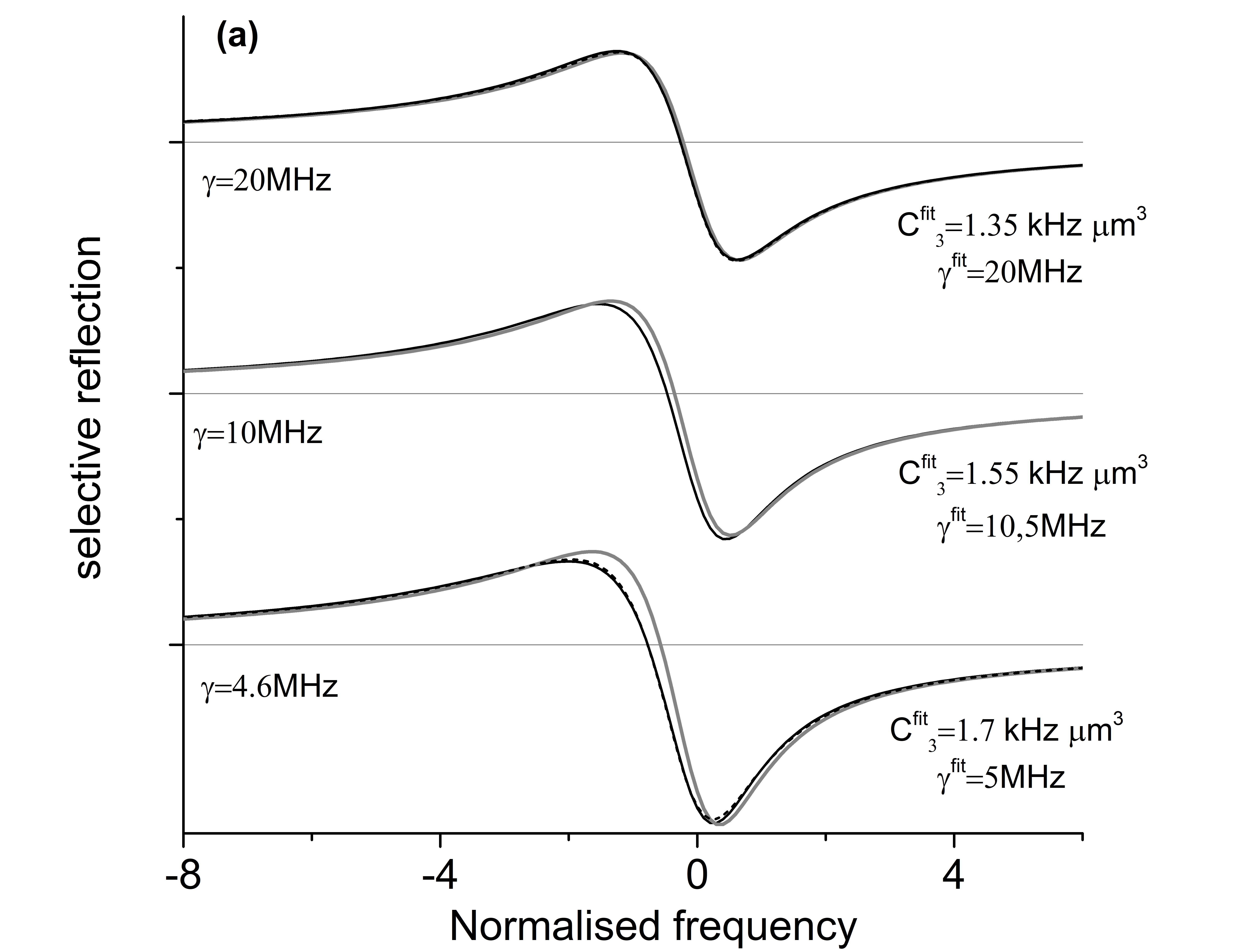}}
\subfigure{\includegraphics[width=84mm]{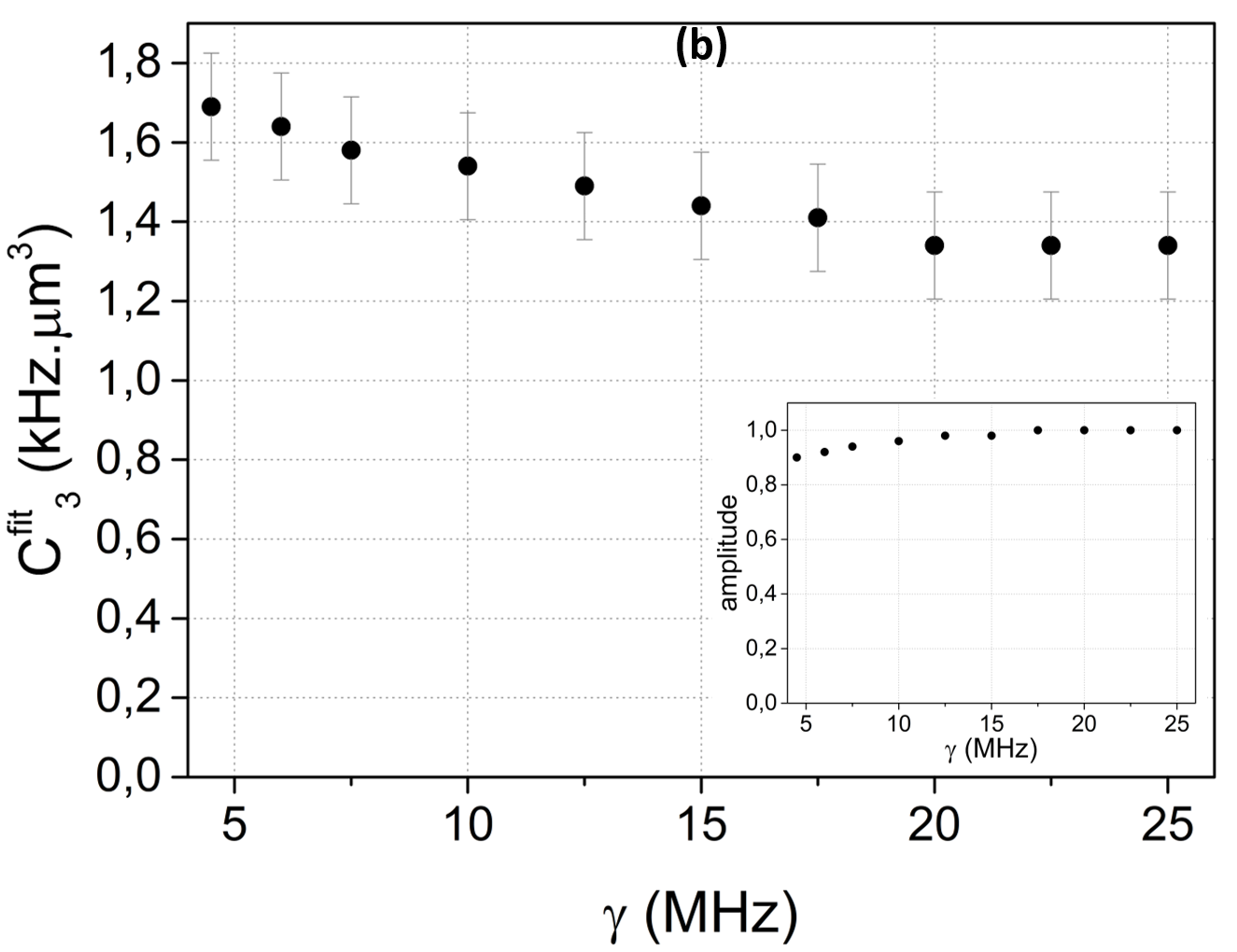}}
\caption{(a) The same as in Fig.\ref{Fig2} but for the $6S_{1/2} \rightarrow 6P_{1/2}$ transition. Black lines are the exact SR spectra, grey lines SR spectra with a pure van der Waals potential and dashed lines are fits of the exact SR spectra using an ad-hoc $C_{3}^{\mathrm{fit}}$ coefficient. The linewidths investigated are $\gamma=4.6\;\mathrm{MHz}$ (natural linewidth) as well as $\gamma=10 \;\mathrm{MHz}$ and $\gamma=20\;\mathrm{MHz}$, giving a $C_{3}^{\mathrm{fit}}$  of $1.7\;\mathrm{kHz} \, \mu \mathrm{m}^{3}$, $1.5\;\mathrm{kHz} \, \mu \mathrm{m}^{3}$ and $1.2\;\mathrm{kHz}\,\mu \mathrm{m}^{3}$  respectively. The theoretical values of the van der Waals coefficient is $C_{3}=1.1\;\mathrm{kHz}\,\mu\mathrm{m}^3$.(b) $C_{3}^{\mathrm{fit}}$ coefficient as a function of linewidth. As {in Fig.\ref{Fig2}}, the inset shows the amplitude ratio between the exact SR spectra and the corresponding fits.
\label{Fig3}}
\end{figure}
     
We now use  the theory developed in the previous section to calculate SR spectra, of the electric quadrupole transition  $6S_{1/2}\rightarrow5D_{5/2}$ . In Fig.\ref{Fig2}(a) we show the calculated SR spectra as black solid lines for different values of the collisional broadening. The grey lines show the expected SR spectra, assuming a pure van der Waals non-retarded law $C_3 z^{-3}$. The differences between spectra are significant, especially for $\gamma=120\;\mathrm{kHz}$ (natural transition linewidth) where differences are indeed striking. This confirms that retardation effects can play a important role in this experiment. To strengthen our analysis we try to fit the fully retarded SR spectra using an ad hoc van der Waals coefficient, $C_3^{\mathrm{fit}}$. The fitting methods have been detailed in numerous works (see for example \cite{papageorgiou_laserphyics_1994, failacheEPJD2003}). We briefly remind that the fitting process optimizes a dimensionless parameter $A=\frac{2C_3 k^3}{\gamma}$, the transition linewidth, and accounts for the amplitude of the spectra as well as a small (pressure induced) shift of the transition frequency. The best fits are shown with dashed lines in Fig.\ref{Fig2}(a), whereas the values of  $C_3^{\mathrm{fit}}$ as a function of transition linewidth are shown in Fig.\ref{Fig2}(b). It is evident that an ad hoc van der Waals model can in most cases satisfactorily fit the fully retarded spectra. It should nevertheless be noted that the quality of the fits clearly degrades as the linewidth decreases (for $\gamma=120\;\mathrm{kHz}$ the fit cannot reproduce very well the retarded SR spectrum). Most importantly, the $C_3^{\mathrm{fit}}$ is not constant  but displays a clear dependence on the transition linewidth $\gamma$, as can be seen in Fig.\ref{Fig2}(b). As $\gamma$ increases the ad hoc van der Waals coefficient approaches its theoretical value $C_3^{\mathrm{eff}}(z\rightarrow0)$, whereas for narrow linewidths, SR seems to probe the Casimir-Polder interaction at a finite distance (more than $100\; \mathrm{nm}$ away from the surface when $\gamma=120\;\mathrm{kHz}$). This phenomenon has a rather transparent interpretation: Due to big Casimir-Polder shifts, atoms that are very close to the surface, experience a large detuning parameter that reduces their relative contribution to the SR spectrum. When the transition linewidth increases due to collisional broadening, $\Delta$ decreases, thus enhancing the contribution of atoms that are closer to the surface. In the inset of Fig. \ref{Fig2}(b) we plot the ratio of amplitudes between the exact SR spectra and the van der Waals fits. Here, also we observe a dependence as a function of linewidth. These variations (about $20\%$) are much smaller than the $C_3^{\mathrm{fit}}$ variations. We also stress that the actual experimental amplitude of the spectra would also depend on the atomic vapor density. As such an experimental study of amplitude effects is more challenging.       

The same analysis is repeated for the $\mathrm{D}_1$ transition of Cesium and the results are summarized in Fig.\ref{Fig3}. Here, the retarded SR lineshapes (solid black lines) can be almost exactly reproduced by an ad hoc van der Waals fit (dashed lines). As previously, the values of $C_3^{\mathrm{fit}}$ (Fig.\ref{Fig3}(b)) also decrease with increasing linewidth converging towards the value of $C_3^{\mathrm{fit}}=1.35\;\mathrm{kHz}\,\mu \mathrm{m}^3$. For $\gamma$ close to the natural linewidth ($4.6\;\mathrm{MHz}$), $C_3^{\mathrm{fit}}=1.7\;\mathrm{kHz}\, \mu \mathrm{m}^3$ a value that cannot be justified only by the Casimir-Polder shift (see the inset in Fig.\ref{Fig1}(a)). In this case, in order to account for the observed dependence of $C_3^{\mathrm{fit}}$ as a function of linewidth one has to consider both the distance dependent shift and linewidth. This is corroborated by the fact that the fitting process gives an ad hoc linewidth $\gamma^{\mathrm{fit}}$ which  is slightly larger than the real values (by about $0.5\;\mathrm{MHz}$), an effect also linked to the distance dependent linewidth close to the surface (see Fig.1(b)). Small variations of the amplitude of the fitted curves are also observed and shown in the inset of Fig.\ref{Fig3}(b).

Contrary to the $6S_{1/2}\rightarrow5D_{5/2}$ transition, the $\mathrm{D}_{1}$ line of Cesium is a particularly strong line with a well separated hyperfine structure, but the predicted retardation effects are smaller. Previous experiments \cite{laliotisAPB2008}, conducted for large linewidths ($\gamma>20\;\mathrm{MHz}$) give a value of $1.4\;\mathrm{kHz}\,\mu \mathrm{m}^3$ , with error bars of about $15\%$. These results are in good agreement with the predictions of Fig.\ref{Fig3}(b). A more conclusive experimental demonstation of retardation requires measurements at small linewidths and probably an improvement of the experimental error bars. This regime was not attained in the experiment presented in  \cite{laliotisAPB2008} mainly due to the existence of impurities in the cesium cell that limited the minimum observable linewidth. When comparing experiment to theory it's also worth keeping in mind that the theoretical estimates of the Casimir-Potential are sensitive to the exact knowledge of the transition probabilities of all the relative dipole couplings as well as the dielectric constant of sapphire (see \cite{laliotisnatcommun2014} for a discussion on the error bars of the theoretical predictions). 

Our analysis also gives the possibility to fit experimental data with a fully retarded library of curves (values of $\mathcal{A} $), which would depend on the specific transition and the specific dielectric investigated. In this case the fitting process would adjust for the transition linewidth ($\gamma$) and the dimensionless parameter ($\eta$) which measures the strength of the Casimir-Polder potential (both shift and linewidth) with respect to its theoretical values, assuming that distance dependence is fixed.

\section{Conclusions}

Here we have focused our analysis in spectroscopic experiments performed with atoms in front of infinite plane surfaces. However, our methodology can be easily extended to composite surfaces, assuming that the Casimir-Polder potentials and propagation optics can be correctly evaluated \cite{chevrollier_selective_2001}. The simplest example of a composite surface is the case where alkali adsorbants are deposited on or even react with the surface. This is a common phenomenon in vapour cells filled with alkali atoms that can strongly depend on the nature of the surface  \cite{Bouchiat1999}. In this respect sapphire windows seem to be more favorable, with an additional benefit of allowing much higher temperatures, than glass or calcium fluoride windows \cite{ passeratlaserphysics2014}. Although a theoretical analysis of the problem is challenging it is probable that a simple van der Waals approximation is not sufficient to analyze these effects. 

A more interesting scenario includes the controlled deposition of bi-dimensional materials such as graphene on a dielectric surface. Already, Casimir force measurements have been performed on a composite dielectric-graphene surface \cite{Banishev_PRB_2013} and  theoretical proposals exist for extending such measurements to the Casimir-Polder domain (see for example \cite{Ribeiro&Scheel_PRA_2013, Klimchitskaya_PRA_2014}). Casimir and Casimir-Polder type measurements allow us to get useful information on the dielectric properties of bi-dimensional materials. More importantly, stacking bi-dimensional layers may eventually allow engineering an effective dielectric constant and the plasmon-polariton modes of the surface. Finally, non-trivial geometries, without cylindrical symmetry, such as gratings \cite{benderprx2014} or metamaterials \cite{Chan_ScieAdv_2017} have already been experimentally explored. In the case of the atom-metamaterial interaction, initial selective reflection measurements indicate that retardation effects are important for a correct interpretation of the experiment. 

In conclusion, we have presented the theoretical background that allows us to take into account the effects of Casmir-Polder retardation in spectroscopic experiments of the atom-surface interaction. We have proposed specific experiments, where retardation can have observable effects. Unlike previous retardation measurements \cite{hindscp1993,benderprl2010} the experiments investigated here are sensitive to the difference of energy shifts between ground and excited state atoms, and therefore sensitive to both non-resonant and resonant components of the atom-surface potential. Our analysis shows that although experimental measurements can in most cases be fitted with a simple van der Waals model, such an analysis will yield a linewidth dependent van der Waals coefficient. This is because the probing depth of the experiments increases with decreasing linewidth. Finally we show that our analysis will be useful when dealing with composite, non-trivial surfaces.

J.C. de Aquino Carvalho thanks the Brazilian  program Ci\^encia Sem Fronteiras for financial support of his PhD thesis. A. Laliotis and J.C. de Aquino Carvalho acknowledge discussions with Daniel Bloch that led to an improvement of the manuscript. A. Laliotis and M. Ducloy acknowledge discussions with David Wilkowski.

\appendix
\section{}

By applying the transformation $s=k(z+z^\prime)$ and $t=k(z-z^\prime)$ Eq.(\ref{eq6}) can be written as:

\begin{equation}
I=\frac{1}{k^2}\int_{0}^{\infty}ds \int_{0}^{s}dt \frac{\frac{t}{k} e^{i s}}{L(\frac{s+t}{2k} )-L(\frac{s-t}{2k})}
\label{eqA1}
\end{equation}

after some algebra the integral $I$ can be written as 

\begin{equation}
I=\frac{2}{\gamma k^2}\int_{0}^{\infty}ds e^{is}e^{-\alpha s} \int_{0}^{s}dt \frac{1}{(1-i \Delta) -i \mathcal{A} \Xi(s,t)}
\label{eqA2}
\end{equation}
where $\Xi(s,t)$ is defined as :
\begin{equation}
\Xi(s,t)= \frac{\xi( \frac{s+t}{2k})- \xi( \frac{s-t}{2k})}{t}  
\label{eqA3}
\end{equation}
whereas $\mathcal{A} = \frac{2 \eta k}{\gamma}  $, $\Delta=\frac{2(\omega-\omega_{o})}{\gamma}$ and $\alpha$ is an attenuation coefficient already defined in the main text. Further algebra leads to the following:
\begin{equation}\begin{split}
I=\frac{2}{(1-i \Delta) \gamma k^2} \left[ \int_{0}^{\infty}s e^{is}e^{-\alpha s} ds  \right] \\
+\frac{2}{(1-i \Delta) \gamma k^2} \left[ \int_{0}^{\infty}ds e^{is}e^{-\alpha s}  \int_{0}^{s}dt \frac{i \mathcal{A} \Xi(s,t))}{(1-i \Delta) -i \mathcal{A} \Xi(s,t)} \right]
\label{eqA4}
\end{split}\end{equation}

The first integration can be performed analytically giving the final expression:
\begin{equation}\begin{split}
I=\frac{2}{(1-i \Delta) \gamma k^2} \left[ \frac{1}{(-i+\alpha)^2} \right] \\ 
+\frac{2}{(1-i \Delta) \gamma k^2}  \left[ \int_{0}^{\infty}ds e^{is}e^{-\alpha s}  \int_{0}^{s}dt \frac{i \mathcal{A} \Xi(s,t)}{(1-i \Delta) -i \mathcal{A} \Xi(s,t)} \right]
\label{eqA5}
\end{split}\end{equation}

Solving numerically Eq. (\ref{eqA5}) can be challenging. We find that the introducing the laser field attenuation parameter helps convergence of the integrals without significantly influencing the final results, so long as $\alpha\ll 1$ (typically $\alpha<0.1$ is sufficient). Also, for large values of $s$ ($s\rightarrow\infty$) the last integral in Eq. (\ref{eqA5}) converges to:

\begin{equation}\begin{split}
\int_{0}^{s}dt \frac{i \mathcal{A} \Xi(s,t)}{(1-i \Delta) -i \mathcal{A} \Xi(s,t)}\rightarrow \\ \frac{A_o}{\sqrt{s}}+\frac{B_{o}}{s}\left[ \cos{(s+\phi)}+i \sin{(s+\phi)}\right]+\frac{B_{1}+iB_{2}}{s}
\label{eqA6}
\end{split}\end{equation}

where $A_o$, $B_o$, $B_1$, $B_2$ are constants that depend on the specific problem in question. The approximation of Eq. (\ref{eqA6}) greatly simplifies calculation of the SR integral in the limiting case $\alpha\rightarrow 0$.

\bibliography{biblio}

\end{document}